\documentclass[prd,aps,showpacs,preprintnumbers,amssymb]{revtex4}
\usepackage{graphicx}

\usepackage{epsf}

\def\e3p{$\eta \rightarrow 3 \pi$}

\begin{document}

\title{%
\hfill{\normalsize\vbox{%
\hbox{}
 }}\\
{Semi-leptonic $D_s^+$(1968) decays as a scalar meson probe}}

\author{Amir H. Fariborz $^{\it \bf a}$~\footnote[1]{Email:
 fariboa@sunyit.edu}}

\author{Renata Jora $^{\it \bf b}$~\footnote[2]{Email:
 rjora@ifae.es}}

\author{Joseph Schechter $^{\it \bf c}$~\footnote[3]{Email:
 schechte@physics.syr.edu}}

\author{M. Naeem Shahid $^{\it \bf c}$~\footnote[4]{Email:
 mnshahid@physics.syr.edu}}

\affiliation{$^ {\bf \it a}$ Department of Engineering, Science
 and Mathematics,
 State University of New York Institute of Technology, Utica,
 NY 13504-3050, USA.}

\affiliation{$^ {\bf \it b}$ Grup de Fisica Teorica and IFAE,
Universitat Autonoma de Barcelona, E-08193 Bellaterra (Barcelona),
Spain.}

\affiliation{$^ {\bf \it c}$ Department of Physics,
 Syracuse University, Syracuse, NY 13244-1130, USA,}

\date{\today}

\begin{abstract}

    The unusual multiplet structures associated with the light
    spin zero mesons have recently attracted a good deal of
    theoretical attention. Here we discuss some aspects
    associated with the possibility  of getting new experimental
    information on this topic from semi-leptonic decays of heavy
    charged mesons into an isosinglet scalar or pseudoscalar plus
    leptons.

\end{abstract}


\maketitle

\section{Introduction}

    For the first years after the quark model was accepted it was believed
    that the lightest scalar meson should be a quark-antiquark composite with
    mass value similar to those of the tensor and axial vector mesons. In
    particular, an occasionally discussed light, broad ``sigma meson" was
    not expected to exist. However, more recent work has provided evidence for
    such a particle as well as for other similar light scalars. (Some characteristic
    references are \cite{pdg} - \cite{FJSS11}.)

    In fact there seem to be enough scalar candidates to fill up two different
     nonets. A model including these states, with also two nonets of low lying
     pseudoscalars in order to form chiral multiplets has been studied in some detail;
     \cite{BFMNS01}, \cite{mixing}, \cite{thermo}, \cite{FJS08}, \cite{FJS05}-\cite{FJS09},
     \cite{FJSS11}.
      Note that chiral symmetry is the exact symmetry of QCD with massless
     quarks. Adding  ``soft" quark mass terms results in ``sigma models" which give many
     reasonable low energy predictions. In the models just mentioned one chiral (i.e. containing
     both scalars and pseudoscalars) nonet is supposed to represent states with a
     quark-antiquark substructure while the other nonet is supposed to represent states
     with a two quark - two antiquark substructure. The physical states are suitable
     linear combinations.

     On the experimental side of the subject, information on the light scalars has often
     been extracted from study of pion pion and other scattering processes. Another way is to search for
     scalar resonances explicitly in particle decay processes. Recently, the CLEO collaboration
     has reported \cite{Cleo} good evidence for the scalar $f_0$(980) in the semi-leptonic decay of the
      $D_s^+$(1968) meson. Since there is more phase space available, it may be possible
      to find other scalar iso-singlet states in this and similar semi-leptonic decays of heavy mesons.
       There are also isosinglet pseudoscalar states like the $\eta$
       and $\eta^\prime$(980) which can be studied and in fact have been already reported in the decays of the
        $D_s^+$(1968).

       As a possibly helpful adjunct to future work in this direction we will, in the present paper,  make some
        theoretical estimates of the semi-leptonic decay widths of the
        $D_s^+$(1968) into the four scalar isosinglet states and the four pseudoscalar isosinglet states
        which are predicted in the chiral model mentioned above.

        In section II we discuss the hadronic ``weak currents" which are needed for the calculation. These are
        mathematically given by the so-called Noether currents of the sigma model Lagrangian being employed.
        We work in the approximation where renormalization of these currents from the symmetry limit are neglected.
        This means that there are no arbitrary parameters available to us. Nevertheless there are some subtleties.
        To explain these we build up the model in three stages rather than just writing the final result
        immediately.

        In section III we give a detailed description of the calculation of the partial decay widths from the currents
        discussed in section II. For this purpose we also use information on the scalar and pseudoscalar
        meson masses and mixings obtained in \cite{FJS09}.

A short summary and discussion is given in section IV.

        In Appendix A we briefly discuss the well known $K\ell$3 decay which has the same general structure
        as the semi-leptonic $D_s$ decays.

\section{Hadronic Currents in various linear sigma models}

      These models give the usual "current algebra" results near the
      threshold of pion-pion scattering but also yield some additional
       interesting features away from threshold.

      \subsection{Chiral SU(3) model}

  The usual chiral nonet $M(x)$ realizing the $q \bar q$ structure
  of the pseudoscalar and scalar mesons is schematically
  written with chiral SU(3) indices displayed as:

\begin{equation}
M_a^{\dot{b}} = {\left( q_{bA} \right)}^\dagger \gamma_4 \frac{1 +
\gamma_5}{2} q_ {aA}, \label{M}
\end{equation}

where $a$ and $A$ are respectively flavor and color indices. For clarity,
on the left hand side the undotted index transforms under the left SU(3)
while the dotted index transforms under the right SU(3).
The decomposition in terms of scalar and pseudoscalar fields is:

\begin{equation}
M=S+i\phi.
\label{m}
 \end{equation}

    Using matrix notation (e.g. $M_a^{\dot{b}} \rightarrow M_{a{\dot{b}}}$)
    the Noether vector and axial currents read (see for example Appendix A of \cite{su71}),

\begin{eqnarray}
    V_{\mu}&=&i\phi\stackrel{\leftrightarrow}{\partial_\mu}\phi +
    iS\stackrel{\leftrightarrow}{\partial_\mu}S,
\nonumber \\
A_{\mu}&=&
S\stackrel{\leftrightarrow}{\partial_\mu}\phi -
    \phi\stackrel{\leftrightarrow}{\partial_\mu}S,
\label{su3currents}
\end{eqnarray}

    The axial symmetry breaking is measured by the vacuum value of $S$:
    \begin{equation}
    S= \tilde{S} + <S>,\quad\quad   <S_a^b> =\alpha_a \delta_a^b,
    \label{su3vevs}
    \end{equation}
     where the normalization is $\alpha_1+\alpha_2= F_{\pi} \approx$ 130.4 MeV
     and $\alpha_1+\alpha_3 = F_{K} \approx$
     156.1 MeV. Note that the overall normalization constant for $V_{\mu}$
     gives the correct value for the ordinary electromagnetic current.
     This determines the normalization for the weak currents in the SU(3)$_L$ $\times$
     SU(3)$_R$ symmetry limit. For the vector currents this amounts to an
     implementation of the ``conserved vector current hypothesis" introduced for beta decay
     many years ago \cite{gz56}. Such an approximation is well known not to be as good for the axial current
     case, but may at least furnish an order of magnitude estimate.
     In detail, with the usual SU(3) tensor indices, the currents read:
\begin{eqnarray}
    V_{\mu a}^b&=&i\phi_a^c\stackrel{\leftrightarrow}{\partial_\mu}\phi_c^b +
    i\tilde{S}_a^c\stackrel{\leftrightarrow}{\partial_\mu}\tilde{S}_c^b
    +i(\alpha_a-\alpha_b)\partial_\mu\tilde{S}_a^b,
\nonumber \\
A_{\mu a}^b&=&\tilde{S}_a^c\stackrel{\leftrightarrow}{\partial_\mu}\phi_c^b -
    \phi_a^c\stackrel{\leftrightarrow}{\partial_\mu}\tilde{S}_c^b
    +(\alpha_a+\alpha_b)\partial_\mu\phi
    _a^b,
\label{su3currents}
\end{eqnarray}

     For example, the relevant hadronic current needed to describe the
     semi-leptonic decay $K^+\rightarrow \pi^0$ + $e^+$ + $\nu_e$ is

     \begin{equation}
      V_{\mu 1}^3 = \phi_1^c\stackrel{\leftrightarrow}{\partial_\mu}\phi_c^3
    +i(\alpha_1-\alpha_3)\partial_\mu\tilde{S}_1^3.
    \label{kl3piece}
    \end{equation}

     A relevant application of this formula is given in Appendix A.

    \subsection{SU(3) $M-M^\prime$ model}

         For this model we first introduce another chiral field, $M^{(2)} $
      constructed out of two quarks and two anti-quarks as:

    \begin{equation}
M_a^{(2) \dot{b}} = \epsilon_{acd} \epsilon^{\dot{b} \dot{e} \dot{f}}
 {\left( M^{\dagger} \right)}_{\dot{e}}^c {\left( M^{\dagger}
 \right)}_{\dot{f}}^d.
\label{Sandphi}
\end{equation}

     This object has the form of a ``molecule" made of two $M$'s.
     Alternatively one can schematically make  two quark- two antiquark states
     denoted by $M^{(3)}$ and $M^{(4)}$, from a diquark combined with an anti-diquark
     in two different ways \cite{BFMNS01}.
      One might as well consider the most general linear combination of $M^{(2)}$,
       $M^{(3)}$ and $M^{(4)}$
      as a field representing an object, $M^\prime$ made from two quarks plus two antiquarks.
      $M^\prime$ has the decomposition,

      \begin{equation}
      M^\prime= S^\prime +i\phi^\prime.
      \label{mprimedec}
      \end{equation}

      Then the Noether currents involve the sum of pieces constructed from the
unprimed fields and from the primed fields. The latter take the form,

         \begin{eqnarray}
    V_{\mu a}^{\prime b}&=&i\phi_a^{\prime c}\stackrel{\leftrightarrow}{\partial_\mu}\phi_c^{\prime b} +
    i\tilde{S}_a^{\prime c}\stackrel{\leftrightarrow}{\partial_\mu}\tilde{S}_c^{\prime b}
    +i(\beta_a-\beta_b)\partial_\mu\tilde{S}_a^{\prime b},
\nonumber \\
A_{\mu a}^{\prime b}&=&S_a^{\prime c}\stackrel{\leftrightarrow}{\partial_\mu}\phi_c^{\prime b} -
    \phi_a^{\prime c}\stackrel{\leftrightarrow}{\partial_\mu}\tilde{S}_c^{\prime b}
    +(\beta_a+\beta_b)\partial_\mu\phi
    _a^{\prime b},
\label{primesu3currents}
\end{eqnarray}

wherein,
 \begin{equation}
    S^\prime = \tilde{S}^\prime + <S^\prime>,\quad\quad   <{S^\prime}_a^b> =\beta_a \delta_a^b.
    \label{moresu3vevs}
    \end{equation}

      The total currents are denoted as:

      \begin{eqnarray}
      V_{\mu a}^b(total)&=&V_{\mu a}^{b}+V^{\prime b}_{\mu a}, \nonumber  \\
       A_{\mu a}^b(total)&=&A_{\mu a}^{b}+A_{\mu a}^{\prime b}.
       \label{totalsu3currents}
       \end{eqnarray}

    In contrast to the chiral SU(3) model above, all the primed and
    corresponding unprimed fields mix
    to give physical fields of definite mass. As a simple example, the
  transformation between the physical $\pi^+$ and $\pi^{\prime +}$ fields and the
  original
fields (say $\phi^+$ and $\phi'^+$) is \cite{FJS05}:
\begin{equation}
\left[
\begin{array}{c}  \pi^+ \\
                 \pi'^+
\end{array}
\right]
=
R_\pi^{-1}
\left[
\begin{array}{c}
                        \phi_1^2 \\
                        {\phi'}_1^2
\end{array}
\right]=
\left[
\begin{array}{c c}
                \cos\theta_\pi & - \sin \theta_\pi
\nonumber               \\
\sin \theta_\pi & \cos \theta_\pi
\end{array}
\right]
\left[
\begin{array}{c}
                        \phi_1^2 \\
                        {\phi'}_1^2
\end{array}
\right],
\label{mixingangle}
\end{equation}
which also defines the transformation matrix, $R_\pi$.

The pion decay constant as well as (formally) the decay constant for the much heavier
$\pi(1300)$ particle are defined by the part of the axial current linear in the
fields:
\begin{eqnarray}
A_{\mu 1}^2(total) &=&F_\pi\partial_\mu \pi^+ + F_{\pi'}\partial_\mu
\pi'^+
+\cdots,
\nonumber \\
F_\pi &=&(\alpha_1+\alpha_2) \cos\theta_\pi -
(\beta_1+\beta_2)\sin\theta_\pi,
\nonumber \\
F_{\pi'} &=&(\alpha_1+\alpha_2)\sin\theta_\pi +
(\beta_1+\beta_2)\cos\theta_\pi.
\label{Fpis}
\end{eqnarray}
 The angle $\theta_\pi$ depends on the detailed dynamics. \cite{FJS05}

 In what follows it will be useful for us to specify the mixing matrix for the
 four isoscalar scalar mesons in this model. A basis for these states is given in terms
 of the four component vector $f=$ $(f_a,f_b,f_c,f_d)$ where,

\begin{eqnarray}
f_a&=&\frac{S^1_1+S^2_2}{\sqrt{2}} \hskip .7cm
n{\bar n},
\nonumber  \\
f_b&=&S^3_3 \hskip .7cm s{\bar s},
\nonumber    \\
f_c&=&  \frac{S'^1_1+S'^2_2}{\sqrt{2}}
\hskip .7cm ns{\bar n}{\bar s},
\nonumber   \\
f_d&=& S'^3_3
\hskip .7cm nn{\bar n}{\bar n}.
\label{fourbasis}
\end{eqnarray}

In the above, the quark content is indicated on the right for convenience.
Note that $s$ stands for a strange quark while $n$ stands for a
non-strange quark.
However these basis states are not mass eigenstates. Again,
the detailed dynamics of the model is required to specify this. For typical
values of the model's input parameters (see \cite{FJS09}) the mass eigenstates
 make up a four vector, $F$ = $L_0^{-1}f$ with,

\begin{equation}
(L_o^{-1})  =
\left[
\begin{array}{cccc}
0.601  &  0.199 &  0.600  &    0.489 \\
-0.107   &  0.189 &  0.643  &     -0.735  \\
0.790   &  -0.050&  -0.391  &   -0.470  \\
0.062 &  -0.960 &   0.272  &   -0.019\\
\end{array}
\right]
\label{4by4rots}
\end{equation}

The physical states are identified, with nominal mass values, as
\begin{equation}
F  =
\left[
\begin{array}{c}
f_0(600)   \\
f_0(980)  \\
f_0(1370)  \\
f_0(1800)\\
\end{array}
\right]
\label{fphys}
\end{equation}

    It will also be interesting for us to give the typical result of the model
for the mixing of the four isoscalar pseudoscalars. The analogous basis states are:

\begin{eqnarray}
\eta_a&=&\frac{\phi^1_1+\phi^2_2}{\sqrt{2}} \hskip .7cm
n{\bar n},
\nonumber  \\
\eta_b&=&\phi^3_3 \hskip .7cm s{\bar s},
\nonumber    \\
\eta_c&=&  \frac{\phi'^1_1+\phi'^2_2}{\sqrt{2}}
\hskip .7cm ns{\bar n}{\bar s},
\nonumber   \\
\eta_d&=& \phi'^3_3
\hskip .7cm nn{\bar n}{\bar n}.
\label{etafourbasis}
\end{eqnarray}

For typical
values of the model's input parameters (see \cite{FJS09}) the mass eigenstates
 make up a four component vector, $P$ = $R_0^{-1}\eta$ with,

\begin{equation}
P  =
\left[
\begin{array}{c}
\eta(547)   \\
\eta(958)  \\
\eta(1295)  \\
\eta(1760)\\
\end{array}
\right]
\label{etaphys}
\end{equation}

(These identifications correspond to the
favored scenario discussed in section V of
\cite{FJS09}). The dynamically determined
mixing matrix is then:

\begin{equation}
(R_o^{-1})  =
\left[
\begin{array}{cccc}
-0.675  &  0.661 &  -0.205  &    0.255 \\
0.722   &  0.512 &  -0.363  &     0.291  \\
-0.134   &  -0.546&  -0.519  &   0.644  \\
0.073 &  0.051 &   0.746  &   0.660\\
\end{array}
\right]
\label{mms}
\end{equation}

\subsection{Hybrid $M$ -$M^{\prime}$ model with a heavy flavor}

    As recently discussed in \cite{FJSS11}, the case of three flavors is special in the sense that
    it is the only one in which a two quark-two antiquark field has the correct chiral
    transformation property to mix (in the chiral limit) with $M$. In order to respect
     this property when a heavy meson is included in the Lagrangian, we should demand that "heavy" spin
    zero mesons be made of just one quark and one antiquark. In a linear sigma
    model the kinetic term would then be written as:

    \begin{equation}
    {\cal L} = -\frac{1}{2} Tr^4(\partial_\mu M \partial_\mu M^\dagger)
               -\frac{1}{2} Tr^3(\partial_\mu M^\prime \partial_\mu M^{\prime\dagger}),
      \label{hybridlag}
       \end{equation}

      where the meaning of the superscript on the trace symbol is
      that the first term should be summed over the heavy quark index as
      well as the three light indices. This stands in contrast to the second term
      which is just summed over the three light quark indices pertaining to the
      two quark - two antiquark field $M^\prime$. Since the Noether
      currents are sensitive only to these
      kinetic terms in the model, the vector and axial vector currents with
      flavor indices 1 through 3 in this model are just the same as in
       Eq.(\ref{totalsu3currents}) above. However if either or both flavor
       indices take on the value 4 (referring to the heavy flavor) the current
       will only have contributions from the field $M$. This should be clarified
       by the following example,

\begin{eqnarray}
 V_{\mu 4}^a(total)&=&V_{\mu 4}^a=i\phi_4^c\stackrel{\leftrightarrow}{\partial_\mu}\phi_c^a +
    i S_4^c\stackrel{\leftrightarrow}{\partial_\mu}S_c^a,
    \nonumber \\
 A_{\mu 4}^a(total)&=&A_{\mu 4}^a=S_4^c\stackrel{\leftrightarrow}{\partial_\mu}\phi_c^a -
    \phi_4^c\stackrel{\leftrightarrow}{\partial_\mu}S_c^a.
\label{heavycurrents}
\end{eqnarray}
    Here the unspecified
    indices can run from 1 to 4.
    This equation is correct by construction but does not tell the whole story since
    the connection between the fields above and the physical states involves,
     as in the preceding cases, the details of
    the non-derivative ("potential") terms of the effective Lagrangian.

\section{Different semi-leptonic decay modes of the $D_s^+$(1968)}

    The initial motivation for this work was the recent experimental
discovery \cite{Cleo} of the semileptonic decay mode,

\begin{equation}
D_s^{+}(1968) \rightarrow f_0(980) e^{+} \nu_e,
\label{dsdecay}
\end{equation}

in which the $f_0(980)$ was identified from its two pion decay mode.

    A relevant generalization is to consider other scalar isosinglet candidates than
    the $f_0(980)$. For example the SU(3) $M$ - $M^\prime$ model contains
    four different isoscalar scalars, $F_i$. In addition, there are four different
     isoscalar pseudoscalars in that model, $P_i$. Here we shall calculate
    the predictions of that model for all eight of these decays in the simplest
    approximation. This should provide some useful orientation. In fact there are no
    parameters which have not already been determined in the previous
    treatment \cite{FJS09} of the model.

    The usual weak interaction Lagrangian is,

\begin{equation}
{\cal L}=\frac{g}{2\sqrt{2}}(J_{\mu}^{-}W_{\mu}^{+}+J_{\mu}^{+}W_{\mu}^{-}),
\label{eq:lag}
\end{equation}

wherein,

\begin{eqnarray}
J_{\mu}^{-} & = & i\bar{U}\gamma_{\mu}(1+\gamma_{5})VD + i\bar{\nu}_{e}\gamma_{\mu}(1+\gamma_{5})e,\nonumber \\
J_{\mu}^{+} & = & i\bar{D}\gamma_{\mu}(1+\gamma_{5})V^{\dagger}U + i\bar{e}\gamma_{\mu}(1+\gamma_{5})\nu_e.
\label{eq:curr}
\end{eqnarray}

Here the column vectors of the quark fields take the form:

\begin{equation}
U=\left[\begin{array}{c}
u\\
c\\
t\end{array}\right],\quad\quad D=\left[\begin{array}{c}
d\\
s\\
b\end{array}\right],
\label{eq:udcol}
\end{equation}

and the CKM matrix, $V$ is explicitly,

\begin{equation}
V=\left[\begin{array}{ccc}
V_{ud} & V_{us} & V_{ub}\\
V_{cd} & V_{cs} & V_{cb}\\
V_{td} & V_{ts} & V_{tb}\end{array}\right].
\label{eq:ckm}
\end{equation}

\begin{figure}[htbp]
\centering
\rotatebox{0}
{\includegraphics[width=7cm,clip=true]{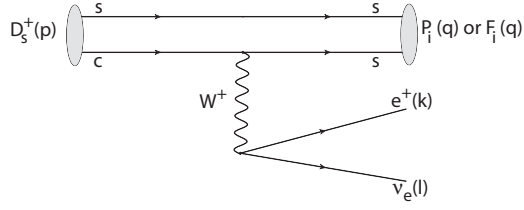}}
\caption[]{
$D_s$ decay.
}
\label{Ds}
\end{figure}

A picture describing the relevant $D_s$ decays is given in Fig. (\ref{Ds}).

The corresponding semi-leptonic decay amplitudes are thus,

\begin{eqnarray}
amp(D_{s}^{+}(p))\rightarrow\left\{ \begin{array}{c}
P_{i}(q)\\
F_{i}(q)\end{array}\right\} +e^+(k)+\nu_e(l))&=&-i\frac{G_{F}}{\sqrt{2}}V_{cs}\left\{ \begin{array}{c}
<P_{i}(q)|V_{\mu4}^{3}(total)|D_{s}^{+}(p)>\\
<F_{i}(q)|A_{\mu4}^{3}(total)|D_{s}^{+}(p)>\end{array}\right\} \nonumber\\
&&\times \bar{u}({\bf l}) \gamma_{\mu}(1+\gamma_{5})v(\bf{k}),
\label{eq:amp}
\end{eqnarray}

where the spinor $v(\bf{k})$ represents the outgoing $e^+$ and $\bar{u}(\bf{l})$ represents the outgoing $\nu_e$.
The relevant hadronic operators can be rewritten in terms of the mass eigenstate scalar isosinglets
and the pseudoscalar isosinglets using Eqs. (\ref{fphys}) and (\ref{etaphys}) as:

\begin{eqnarray}
 V_{\mu 4}^3(total)&=&iD_s^+\stackrel{\leftrightarrow}{\partial_\mu}\phi_3^3 +\cdots
     \nonumber \\
 &=&iD_s^+\sum_{j}(R_0)_{2j}\stackrel{\leftrightarrow}{\partial_\mu}P_j +\cdots
\label{vcomp}
\end{eqnarray}

\begin{eqnarray}
 A_{\mu 4}^3(total)&=&-D_s^+\stackrel{\leftrightarrow}{\partial_\mu}S_3^3 +\cdots
     \nonumber \\
 &=&-D_s^+\sum_{j}(L_0)_{2j}\stackrel{\leftrightarrow}{\partial_\mu}F_j +\cdots
\label{acomp}
\end{eqnarray}

The transposed matrices $L_0$ and $R_0$ are given in Eqs. (\ref{4by4rots})
and (\ref{mms}) respectively, based on a typical numerical solution for
the model parameters \cite{FJS09}. Next the amplitudes are given by,

\begin{equation}
amp(D_{s}^{+}(p)\rightarrow\left\{ \begin{array}{c}
P_{i}(q)\\
F_{i}(q)\end{array}\right\} +e^+(k)+\nu_e(l))=\frac{G_{F}}{\sqrt{2}}V_{cs}\left\{ \begin{array}{c}
(R_0)_{2i}\\
-i(L_0)_{2i}\end{array}\right\}(p_\mu+q_\mu) \bar{u}({\bf l}) \gamma_{\mu}(1+\gamma_{5})v(\bf{k}),
\label{eq:ampeval}
\end{equation}

The squared amplitudes, summed over the emitted lepton's spins, are then,
\begin{equation}
G_F^2|V_{cs}|^2 \frac{1}{m_e^2} \left\{ \begin{array}{c}
((R_0)_{2i})^2\\
((L_0)_{2i})^2
\end{array}\right\}[2k\cdot(p+q){l\cdot(p+q)}-l\cdot k(p+q)^2],
\label{sqamp}
\end{equation}
wherein $m_e$ has been set to zero except for the overall $1/m_e^2$ factor.

This yields the unintegrated decay width,

\begin{equation}
\frac{d\Gamma}{d|\bf{q}|}= \frac{G_F^2|V_{cs}|^2}{12\pi^3} \left\{ \begin{array}{c}
((R_0)_{2i})^2\\
((L_0)_{2i})^2
\end{array}\right\} m(D_s)\frac{|{\bf q}|^4}{q_0}.
\label{udw}
\end{equation}

For integrating this expression we need,

\begin{equation}
|q_{max}|=\frac{m^2(D_s)-m^2_i}{2m(D_s)},
\label{qmax}
\end{equation}
where $m_i$ is the mass of the isosinglet meson $F_i$ or $P_i$ and also the indefinite integral formula, where $x=|\bf{q}|$,

\begin{equation}
\int \frac{x^4dx}{\sqrt{x^2+m_i^2}}=\frac{x^3}{4}\sqrt{x^2+m_i^2}-\frac{3}{8}m^2_i x\sqrt{x^2+m_i^2}+
\frac{3}{8}m^4_i ln(x+\sqrt{x^2+m_i^2}).
\label{integral}
\end{equation}

Table \ref{pscalars} summarizes the calculations of the predicted widths, for $D_s^+$
decays into the four pseudoscalar singlet mesons ($\eta_1=\eta$(547), $\eta_2=\eta$(982),
$\eta_3=\eta$(1225), $\eta_4=\eta$(1794). Notice that the listed masses, $m_i$ are the ``predicted" ones
in the present model) and leptons.

\begin{table}[htbp]
\begin{center}
\begin{tabular}{c||c|c||c}
\hline \hline
$m_i$ (MeV) & $(R_0)_{2i}$ & $(q_{max})_i$ (MeV) & $\Gamma_i$ (MeV)
\\ \hline
553 & 0.661 & 906.20 & 4.14 $\times$ 10$^{-11}$
\\ \hline
982 & 0.512 & 739.00 & 7.16 $\times$ 10$^{-12}$
\\ \hline
1225 & -0.546 & 602.74 & 2.57 $\times$ 10$^{-12}$
\\ \hline
1794 & 0.051 & 166.31 & 2.65 $\times$ 10$^{-17}$
\\ \hline
\hline
\end{tabular}
\end{center}
\caption[]{pseudoscalars.}
\label{pscalars}
\end{table}

Table \ref{scalars}, with the same conventions, summarizes the calculations of the predicted widths for $D_s^+$
decays into the four scalar singlet mesons (${(f_1,f_2,\cdots)=(\sigma,f_0(980),\cdots)}$) and leptons.

\begin{table}[htbp]
\begin{center}
\begin{tabular}{c||c|c||c}
\hline \hline
$m_i$ (MeV) & $(L_0)_{2i}$ & $(q_{max})_i$ (MeV) & $\Gamma_i$ (MeV)
\\ \hline
477 & 0.199 & 933.23 & 4.56 $\times$ 10$^{-12}$
\\ \hline
1037 & 0.189 & 710.79 & 7.80 $\times$ 10$^{-13}$
\\ \hline
1127 & -0.050 & 661.30 & 3.62 $\times$ 10$^{-14}$
\\ \hline
1735 & -0.960 & 219.21 & 3.85 $\times$ 10$^{-14}$
\\ \hline
\hline
\end{tabular}
\end{center}
\caption[]{scalars.}
\label{scalars}
\end{table}

Experimental data exist for only three of these eight decay modes.

\begin{eqnarray}
\Gamma (D_s^+\rightarrow \eta e^+ \nu_e)&=&(3.5 \pm 0.6) \times 10^{-11}\quad \rm{MeV}\nonumber\\
\Gamma (D_s^+\rightarrow \eta^\prime e^+ \nu_e)&=&(1.29 \pm 0.30) \times 10^{-11}\quad \rm{MeV}\nonumber\\
\Gamma (D_s^+\rightarrow f_0 e^+ \nu_e)&=&(2.6 \pm 0.4) \times 10^{-12}\quad \rm{MeV}
\label{expw}
\end{eqnarray}
    It is encouraging that even though our calculation utilized the simplest model
    for the current and no arbitrary parameters were introduced, the prediction for the
    lightest hadronic mode,
    $\Gamma (D_s^+\rightarrow \eta e^+ \nu_e)$ agrees with the measured value. In the case of
    the decay $D_s^+\rightarrow \eta e^+ \nu_e$ the predicted width is about 30$\%$ less than the
    measured value. For the mode $D_s^+\rightarrow f_0(980) e^+ \nu_e$ our predicted value is about
    one third the measured value. Conceivably, considering the large predicted width into the very broad
    sigma state centered at 477 MeV, some of the higher mass sigma events might have been counted as $f_0$(980)
     events, which would improve the agreement. It would be very interesting to obtain
     experimental information about
     the energy regions relevant to the other five predicted isosinglet modes.

\begin{figure}[t]
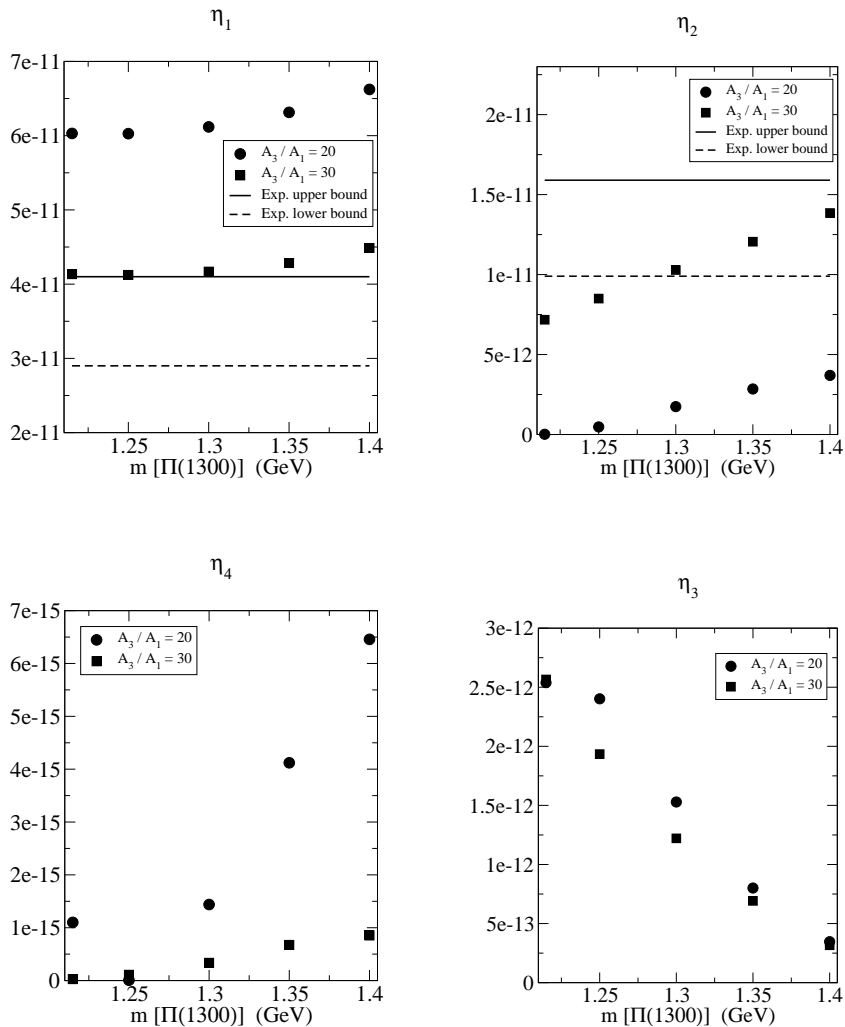

\begin{center}
\epsfxsize = 5cm
\epsfbox{eta1b.eps}
\hskip 1cm
\epsfxsize = 5cm
\epsfbox{eta2b.eps}\\

\vskip 1cm

\epsfxsize = 5cm
\epsfbox{eta4b.eps}
\hskip 1cm
\epsfxsize = 5cm
\epsfbox{eta3b.eps}
\end{center}
\caption[]{%
Starting from the upper left and proceeding clockwise:
The dependences of the pseudoscalar partial widths on the current quark mass ratio $A_3/A_1$ and on
the value of the $\Pi$(1300) mass.
 }
\label{etas}
\end{figure}

\begin{figure}[t]
\begin{center}
\epsfxsize = 5cm
\epsfbox{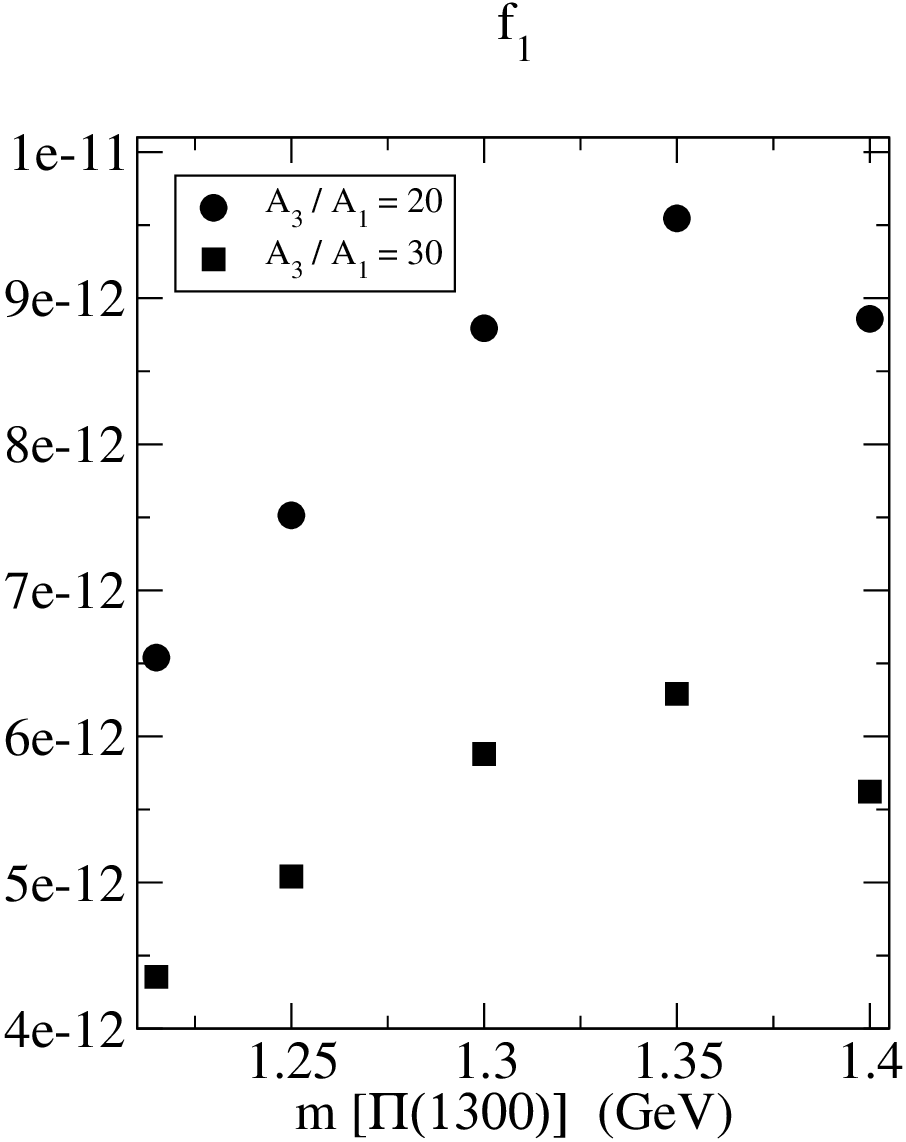}
\hskip 1cm
\epsfxsize = 5cm
\epsfbox{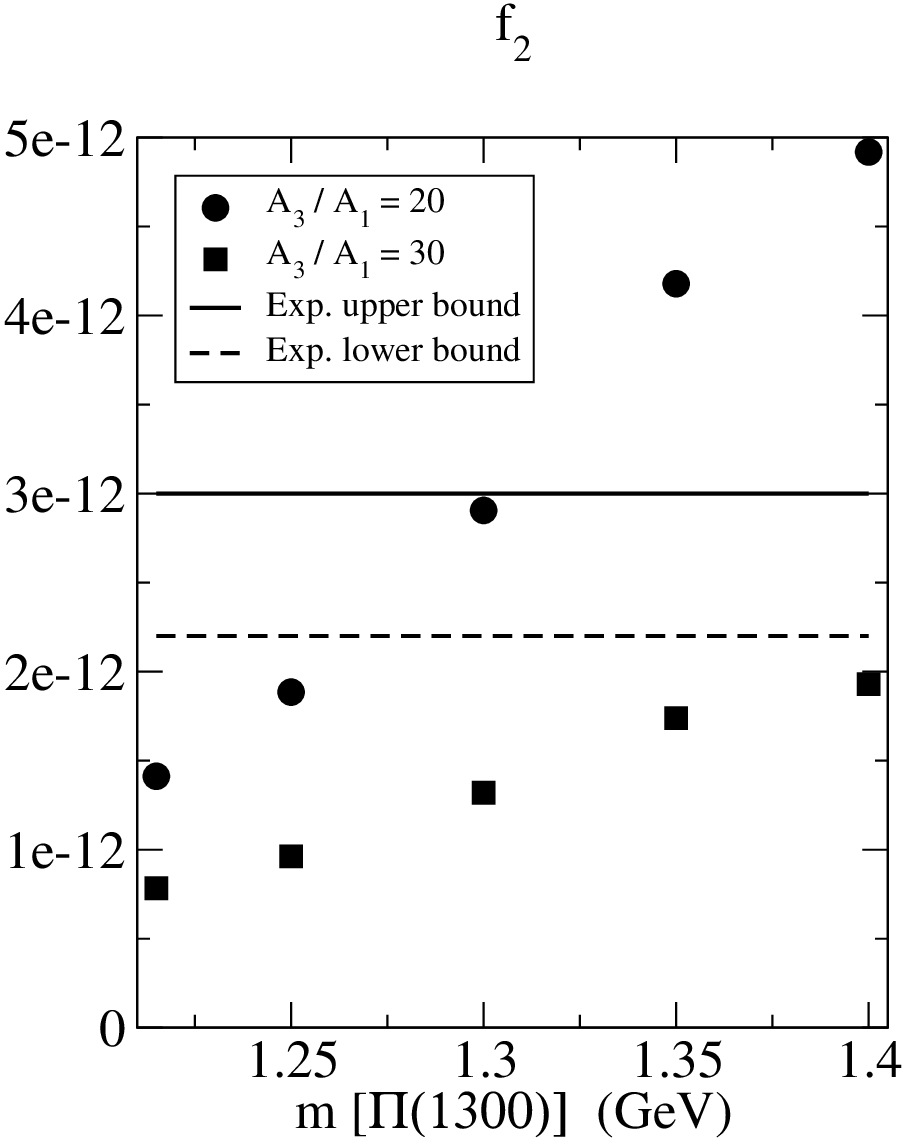}\\

\vskip 1cm

\epsfxsize = 5cm
\epsfbox{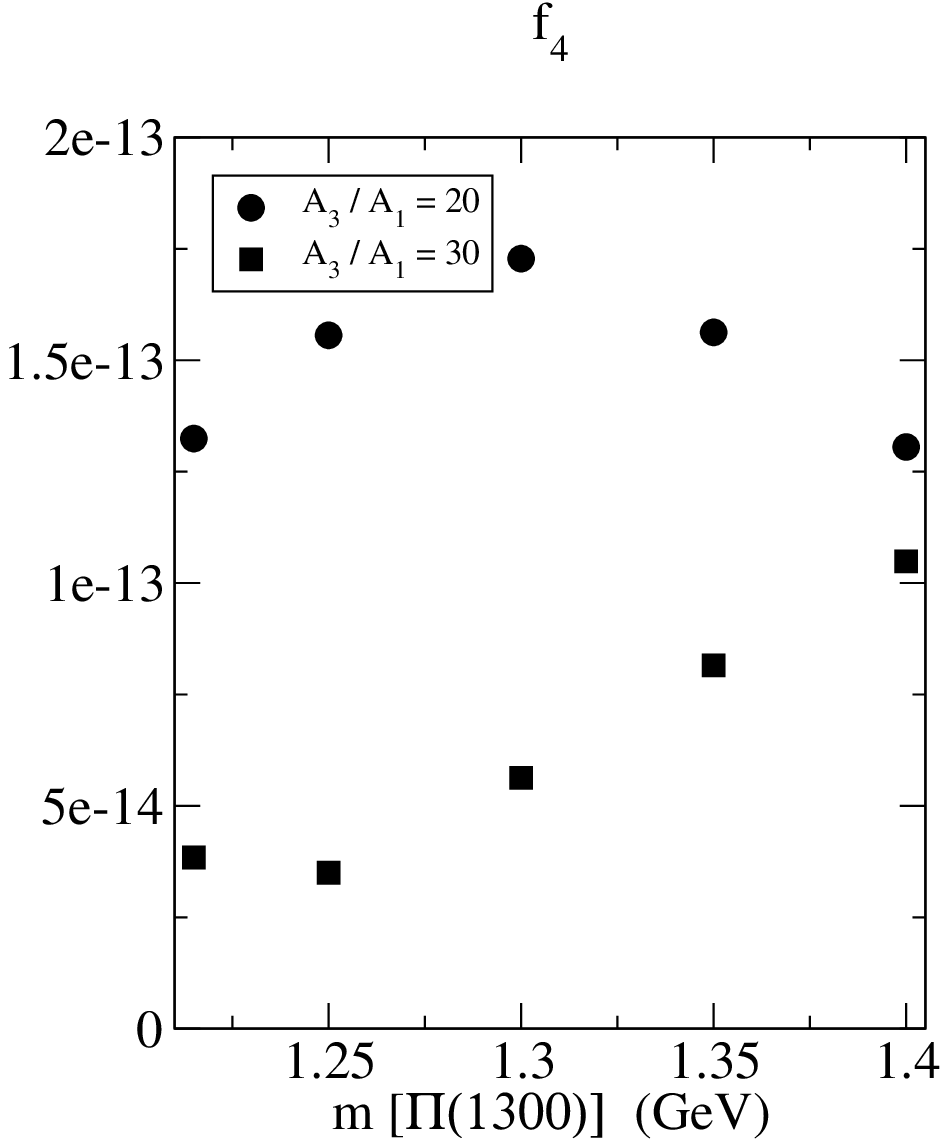}
\hskip 1cm
\epsfxsize = 5cm
\epsfbox{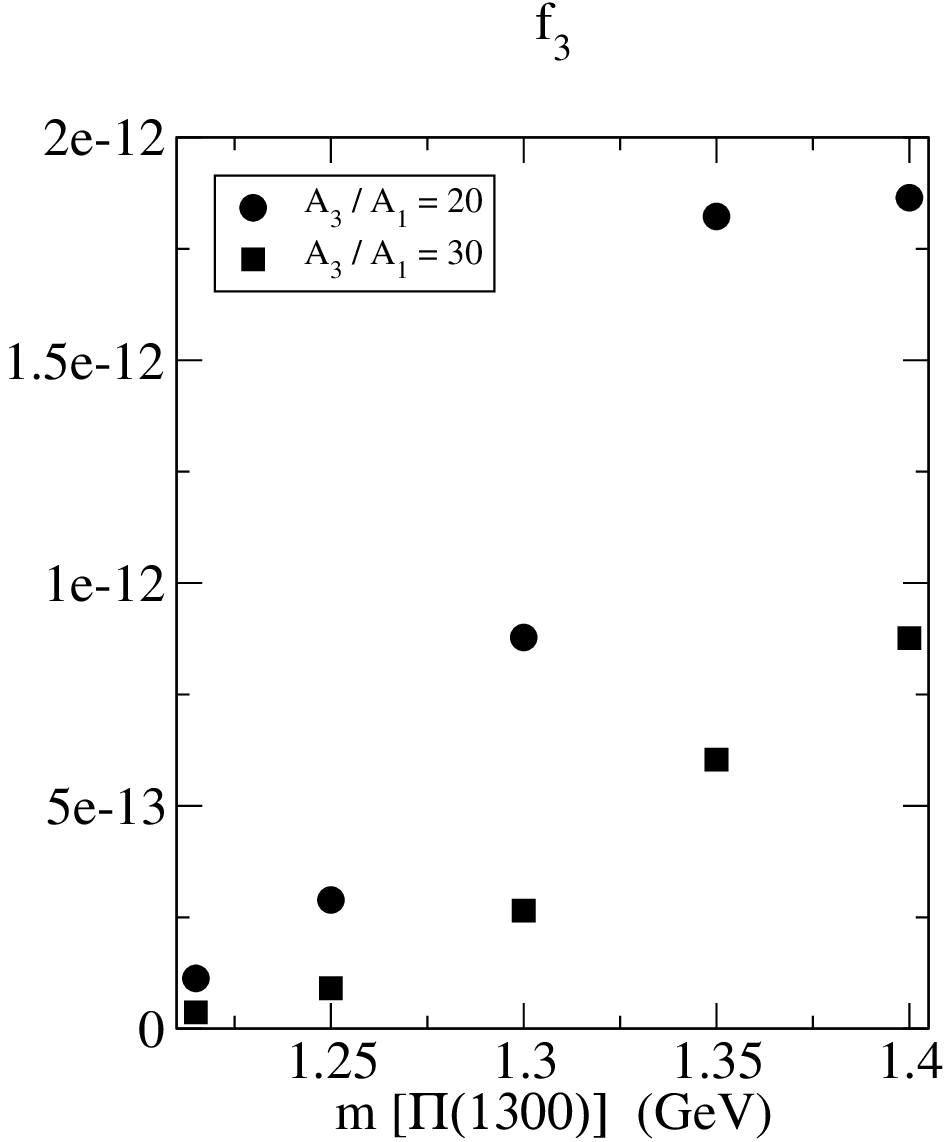}
\end{center}
\caption[]{%
Starting from the upper left and proceeding clockwise:
The dependences of the scalar partial widths on the current quark mass ratio $A_3/A_1$ and on
the value of the $\Pi$(1300) mass.}
\label{fs}
\end{figure}

     Furthermore, these width predictions are based on Eqs. (\ref{4by4rots}) and (\ref{mms}) corresponding to
     particular choices for the quark mass ratio $A_3/A_1$ and the precise mass of the very broad $\Pi$(1300)
     resonance. Varying these within the allowable ranges gives rise to the allowed range of predictions
     displayed in Figs. (\ref{etas})  and (\ref{fs}). One can see that raising m[$\Pi$(1300)]
     and/or lowering $A_3/A_1$ yields better agreement for the predicted semi-leptonic decay width
     of the $f_0$(980). Clearly, the simple model here provides reasonable estimates for
     the semileptonic decay widths of the $D_s^+$(1968).

\section{Summary and discussion}

We saw that the partial widths for semi-leptonic decays of the $D_s^+$(1968)
into isoscalar scalar singlets and pseudoscalar singlets plus leptons could be well estimated in a simple model
where the hadronic current was taken to be the Noether current associated with a minimal linear
sigma model.

The agreement between experiment and theory was better for the decays into the $\eta$ and $\eta^\prime$
than for the decay into the $f_0$(980). The former involve the hadronic vector current, which is ``protected"
according to the conserved vector current hypothesis, while the latter involves the ``unprotected"
axial vector current.

Clearly it would be interesting to try this technique for other semi-leptonic decays of charmed mesons and
 also for bottom mesons. We considered the case when the charged lepton was $e^+$ rather than the cases
 of $\mu^+$ or $\tau^+$. In those two cases an additional form factor as in the calculation of the $K\ell$3
 decay discussed in Appendix A should be taken into account.

Information about the scalars, involving however more work for disentangling the effects of the
strong interaction, can also be obtained from the non-leptonic decay modes of the charm and bottom
mesons. A treatment of $D_S^+ \rightarrow f_0(980)+\pi^+$ has already been carried out \cite{eldl}. The study of the decay like
$B_c^+ \rightarrow scalar + e^+ + \nu_e$ might be useful for learning about mixing between
a $c\bar{c}$ scalar and the lighter three flavor scalars.

 A straightforward, but not necessarily short, improvement of this calculation would be to include
 both vector and axial vector mesons in the starting Lagrangian from which the currents are calculated.

 \section*{Acknowledgments} \vskip -.5cm
 We would like to thank S. Stone for pointing out and discussing the
 CLEO results \cite{Cleo}.
The work of
A.H.Fariborz has been partially supported by the NSF
Grant 0854863 and by a 2011 grant from the office of the Provost of SUNYIT.
The work
of R.Jora has been supported by
CICYT-FEDEF-FPA 2008-01430. The work of
 J.Schechter and M.N. Shahid was supported in part by the U.
S. DOE under Contract no. DE-FG-02-85ER 40231.

\appendix
\section{$K \ell 3$ decay}

     As an illustration of Eq. (\ref{kl3piece}) we consider the matrix
     element, between an initial $K^+$ state with 4-momentum $k$
      and a final $\pi^0$ state with four momentum $p$, of the strangeness
      changing vector current $V_{\mu 1}^3$,

      \begin{equation}
<\pi^0(p)|V_{\mu 1}^3|K^+(k)> \sim f_+(t)(k+p)_\mu + f_-(t)(k-p)_\mu,
       \label{fplusminus}
        \end{equation}

        where $t$ = -$(k-p)^2$.

\begin{figure}[htbp]
\centering
\rotatebox{0}
{\includegraphics[width=7cm,clip=true]{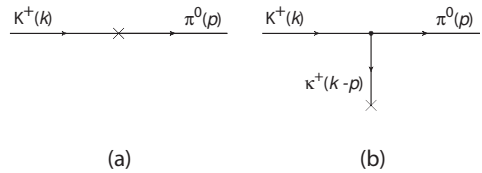}}
\caption[]{
$K\ell3$ decay hadronic current.
}
\label{kdiagram}
\end{figure}

        The first term of Eq. (\ref{kl3piece}) contributes at tree level to the $f_+$
        form factor while the second term contributes to the $f_-$ form
        factor.  These two contributions are illustrated in
        Figs. \ref{kdiagram}a and \ref{kdiagram}b in which the W boson
         which is connected to the leptonic current acts at the points X.
          Here we are evaluating this matrix element
        in the framework of the plain SU(3) linear sigma model in which, furthermore,
        the vector and axial vector  mesons have not been included.

          According to the usual Feynman rules,

         \begin{eqnarray}
         f_+ &=& -\frac{1}{\sqrt{2}},  \nonumber \\
         f_- &=& -\frac{1}{\sqrt{2}} [\frac{\alpha_3-\alpha_1}{\alpha_3+\alpha_1}]
         [\frac{m_\kappa^2-m_\pi^2}{m_\kappa^2-t}],
         \label{kl3formfactor}
         \end{eqnarray}

          wherein $m_\kappa$ denotes the mass of the strange scalar particle contained in
          this model. Furthermore, the explicit form of the $K\kappa\pi$ coupling
          constant in the model was used
          in the expression for $f_-$  \cite{BFMNS01}. Notice that the first bracket in the equation for
          $f_-$ evaluates to about 0.16 and that the physical kappa mass is about 800 MeV in the plain
          SU(3) linear sigma model.

              It is interesting that this decay allows one to learn something about the properties of the
              kappa meson. For this purpose it is necessary to use the process where a final $\mu^+$
              is observed rather than a final $e^+$. That is because the contribution of $f_-(t)$
              to the decay width is proportional to the final lepton mass. Of course the effect of
              the $K^*$(892), which contributes importantly to the $f_+(t)$ form factor should also be included
              to get increased accuracy.

\end{document}